\documentclass[prl,aps,twocolumn,superscriptaddress,showpacs]{revtex4}
\usepackage{graphicx,rotating,subfigure,amsmath,amsfonts,amssymb,delarray}
\renewcommand{\vec}[1]{\boldsymbol #1}

\begin{document}
\bibliographystyle{apsrev}


\title{Entropy Driven Dimerization in a One-Dimensional Spin-Orbital Model}



\author{J. Sirker}
\affiliation{Theoretische Physik I, Universit\"at Dortmund, Otto-Hahn-Str.\!\! 4, D-44221 Dortmund, Germany}

\author{G. Khaliullin}
\affiliation{Max-Planck-Institut f\"ur Festk\"orperforschung, Heisenbergstrasse 1, D-70569 Stuttgart, Germany}


\date{\today}

\begin{abstract}

We study a new version of the one-dimensional
spin-orbital model with spins $S=1$ relevant to cubic vanadates.
At small Hund's coupling $J_H$ we discover dimerization in a pure
electronic system solely due to a dynamical spin-orbital
coupling. Above a critical value $J_H$, a uniform ferromagnetic 
state is stabilized at zero temperature.  
More surprisingly, we observe a temperature driven dimerization 
of the ferrochain, which occurs due to a large entropy released 
by dimer states. This dynamical dimerization seems to be the mechanism 
driving the peculiar intermediate phase of YVO$_3$.     

\end{abstract}
\pacs{75.10.Jm, 75.40.Mg, 05.70.-a} 

\maketitle


In transition metal oxides with perovskite structure the $d$-orbital level
splits into a triplet of $t_{2g}$ symmetry and 
an $e_g$ doublet at higher energy. 
The sign and magnitude of the spin-spin interactions is determined by the 
orbital occupation resulting in a strong spin-orbital coupling. 
This leads to unusual magnetic properties 
of many vanadium, titanium and manganese oxides \cite{TokuraNagaosa}.
The simplest spin-orbital model for spin-1/2 objects with twofold degenerated 
orbitals is the SU(4) symmetric model, where Hund's coupling, 
orbital anisotropy as well as the Jahn-Teller effect 
are ignored \cite{LiMa}. The one-dimensional version of
this model 
has been solved by Bethe ansatz \cite{Sutherland}. 
The ground state and 
the excitation spectrum consisting
of three gapless mixed spin-orbital modes is obtained. The idea of mixed
spin-orbital excitations was also crucial 
to understand anomalous magnetic properties of
LaTiO$_3$ \cite{Keimer}, a system 
showing a coherent orbital-liquid 
ground state \cite{KhaliullinMaekawa}. Thermodynamic properties 
of the SU(4) symmetric model in 1D have been investigated by the quantum Monte Carlo method \cite{FrischmuthMila}. 

Recently vanadium oxides, where the $t_{2g}$ orbitals are filled by two
electrons, have attracted much attention, because of their 
unusual magnetic properties including 
temperature-induced magnetization reversals in YVO$_3$~\cite{RenPalstra2}. 
YVO$_3$ shows a very puzzling 
intermediate phase at $77 < T < 118 K$ with $C$-type magnetic 
order (ferromagnetic 
chains along the $c$ axis with 
antiferromagnetic coupling between the chains), which changes
to a conventional $G$-type N\'eel structure at $T = 77 K$ through a
first order transition 
\cite{Kaw94}.
A significance of $t_{2g}$ orbital degrees of 
freedom for understanding the peculiar magnetic behavior  
of YVO$_3$ has been emphasized \cite{
RenPalstra,NoguchiNakazawa,Bla01,KhaliullinHorsch}.

The relevant spin-orbital model to describe magnetism of vanadium oxides 
has to involve $S=1$ spins because Hund's coupling $J_H$ is large 
\cite{Mila00}.
For the cubic vanadates LaVO$_3$ and YVO$_3$ a realistic 
superexchange model has been derived in Ref.
\cite{KhaliullinHorsch} where two out of the three $t_{2g}$ orbitals are
active along a given cubic axis leading to a $\tau=1/2$ orbital pseudospin. In
the 
classical limit $S\gg 1$ and neglecting Hund's coupling Shen
{\it et al.}~\cite{ShenXie} have shown that the ground state in 1D is 
an orbital valence bond (OVB) solid formed by orbital
singlets and parallel spins, where neighboring OVB's are noninteracting. 

In this paper we treat a realistic quantum spin-orbital model with $S=1$ and
a twofold orbital degeneracy in 1D where 
the effects due to Hund's
coupling are included. Thermodynamics of 
two distinct phases,
a 4-site periodic quantum dimer phase, and a 
ferromagnetic state that occur at larger $J_H$, 
is studied numerically.
A critical value of $J_H$ separating 
the two phases is obtained.
The most important observation is that 
strong dimer correlations
develop {\it at finite temperature on the ferromagnetic side} 
of the transition, which are driven by large entropy of
low-lying dimer states of the model. We discuss the relevance 
of these findings to the intermediate phase of YVO$_3$.

The Hamiltonian of the model is given by
\begin{equation}
\label{eq1}
H = J \sum_i \left[\frac{1}{2} \left(\vec{S}_i\cdot \vec{S}_{i+1} + 
1 \right) \hat{J}_{i,i+1} + \hat{K}_{i,i+1} \right] \; ,
\end{equation}
where $\vec{S}$ being an $S=1$ spin operator, and $J=4t^2/U$ represents
an overall superexchange energy scale.
The operators $\hat{J}_{i,j}$ and $\hat{K}_{i,j}$ describe 
orbital exchange processes on the bond
\begin{eqnarray}
\label{eq2}
%
\hat{J}_{i,j} &=& (1+2R)\left(\vec{\tau}_i\cdot 
\vec{\tau}_j + \frac{1}{4}\right) 
- r \left(\tau_i^z\tau_j^z + \frac{1}{4}\right) - R \; , \nonumber \\
\hat{K}_{i,j} &=& R \left(\vec{\tau}_i\cdot\vec{\tau}_j + 
\frac{1}{4}\right) + r \left(\tau_i^z\tau_j^z + \frac{1}{4}\right) \; ,
\end{eqnarray}   
with $\vec{\tau}$ acting in a $\tau=1/2$ orbital pseudospin space. 
The coefficients  $R=\eta/(1-3\eta)$ and $r=\eta/(1+2\eta)$ originate from
the Hund's rule splitting of the excited $t_{2g}$ multiplet by $\eta=J_H/U$. 
This is 
the Hamiltonian derived for cubic vanadates
in Ref.~\cite{KhaliullinHorsch} restricted here to one cubic axis.
Physically, pseudospin $\vec{\tau}$ describes 1D fluctuations of $xz$
and $yz$ orbitals along the $c$ axis, while the third orbital $xy$ is
frozen below a structural transition at $\sim 200 K$
\cite{RenPalstra,NoguchiNakazawa,Bla01}. 

To study the thermodynamic properties we apply the transfer-matrix DMRG, a numerical method
which is particularly suited because the thermodynamic limit is performed
exactly.
By means of the Trotter formula the partition function is expressed 
as a product of local
transfer matrices evolving along the spatial direction. For the calculations
we use the well known decomposition of the Hamiltonian into even and odd parts
leading to a classical model with checkerboard structure \cite{Peschel}
as well as a different Trotter-Suzuki mapping leading to a classical model
with alternating rows \cite{SirkerKluemper2}. Within the alternative mapping
correlation lengths (CLs) and corresponding wavevectors are determined
unambiguously by the eigenvalues of the quantum transfer matrix (QTM). In the
thermodynamic limit the free energy in both cases is given solely by the
largest eigenvalue of the QTM, which is real and unique. The infinite DMRG
algorithm is used to decrease the temperature where up to 120 states are kept
in each renormalization step.

First, we consider the Hamiltonian for $\eta=0$ (referred to as ``isotropic''
point below) where Eq.~(\ref{eq1}) simplifies to 
\begin{equation}
\label{eq4}
H_0 = \frac{J}{2}\sum_i \left(\vec{S}_i\cdot \vec{S}_{i+1} + 1 \right)\left(\vec{\tau}_i\cdot\vec{\tau}_{i+1} + \frac{1}{4}\right).
\end{equation}
Note, that the model has a SU(2)$\times$SU(2)
symmetry but not the higher SU(4) symmetry group as possible in the $S=1/2$ case
\cite{LiMa}. 
Regarding only a single bond, the lowest energy $-J/2$ is obtained if the
orbitals form a singlet with $\langle\vec{\tau}_i\cdot\vec{\tau}_{i+1}\rangle=-3/4$ 
and the spins are ferromagnetically aligned. The first excited state is given
by a spin singlet/orbital triplet and is separated by a gap of $J/4$ only.
Therefore a strong orbital dimerization in the ground state is expected leading
to alternating ferromagnetic-antiferromagnetic spin exchange couplings, which would be
given by $-J/4$ and $+J/8$ for perfect dimerization. Due to translational
invariance the corresponding classical ground state would be fourfold
degenerate. We also might expect mixed spin-orbital excitations playing an
important role in thermodynamics.
\begin{figure}[!ht]
\includegraphics*[width=0.9\columnwidth]{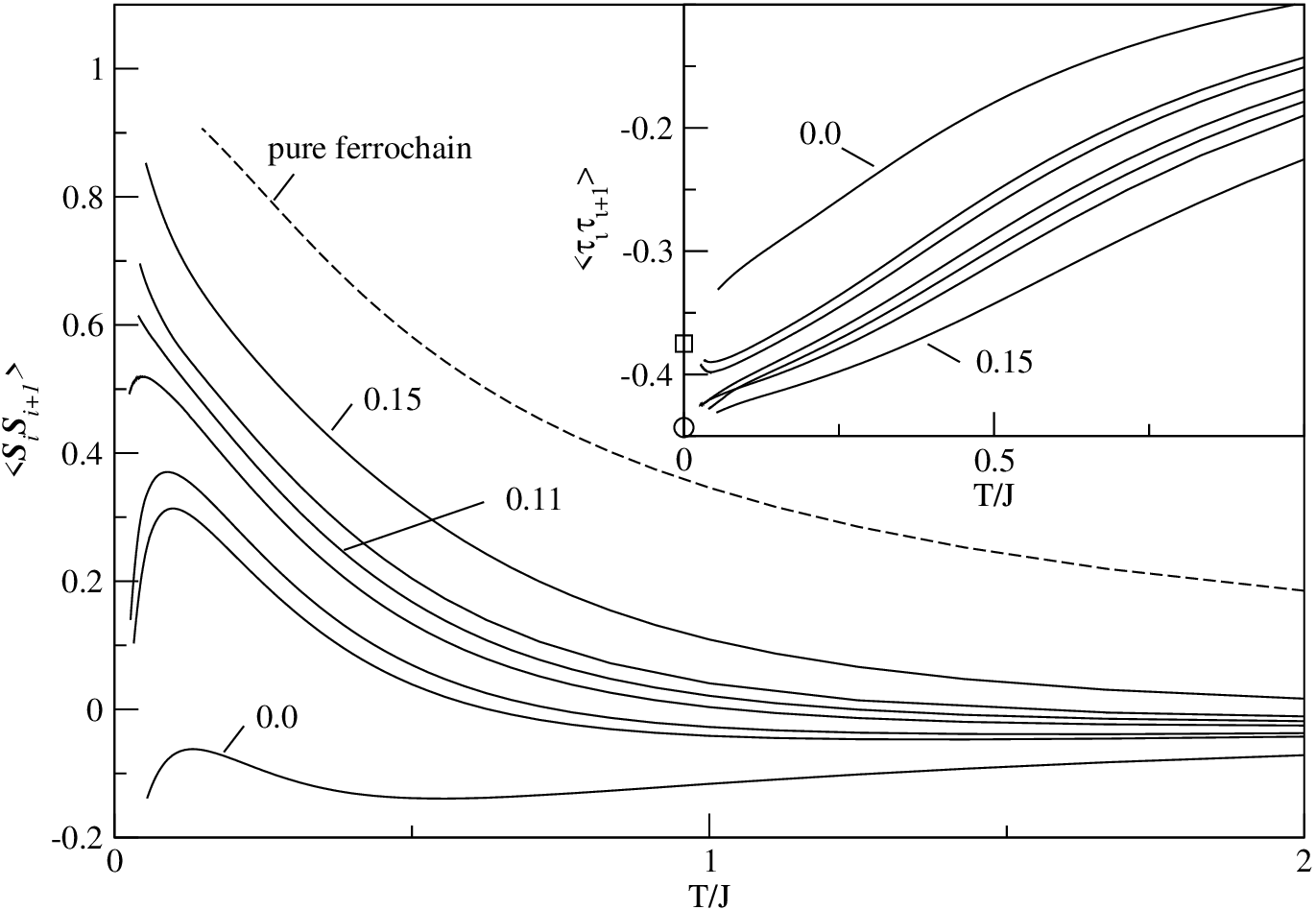}
\caption{The main figure (inset) shows the temperature dependence of nearest
  neighbor correlations $\langle\vec{S}_i \cdot \vec{S}_{i+1}\rangle$
  ($\langle\vec{\tau}_i \cdot \vec{\tau}_{i+1}\rangle$) for $\eta$
  values 0.0, 0.07, 0.08, 0.1, 0.11, 0.12, 0.15. For comparison the spin only case is shown where $J_{i,j}$ is
  fixed by its zero temperature expectation value for $\eta=0.15$. The circle (square) at
  $T=0$ in the inset denotes $-\ln 2 +1/4$ ($-3/8$).}
\label{fig1}
\end{figure}
In Fig.~\ref{fig1} the temperature dependence of the nearest neighbor spin-spin correlation function $\langle \vec{S}_i\cdot \vec{S}_{i+1}\rangle$ and orbital-orbital correlation function $\langle \vec{\tau}_i\cdot \vec{\tau}_{i+1}\rangle$
for different $\eta$ values is shown. First, we want to discuss in more detail the case
$\eta=0$ and turn to finite Hund's coupling later on. Note, that in the zero temperature limit the nearest neighbor
orbital correlation approaches 
$-3/8$, the value for perfect dimerization. In the high temperature limit the orbitals are completely disordered and the spin pair correlation is therefore negative. While lowering the temperature orbital singlets are formed and $\langle \vec{S}_i\cdot \vec{S}_{i+1}\rangle$ increases due to the ferromagnetic spin interaction within an orbital singlet. The curve reaches a maximum  at $T/J\approx 0.13$ corresponding to
the energy scale of the orbital gap (see below),
and antiferromagnetic quantum fluctuations of spins in neighboring orbital singlets are strongly enhanced and dominate below. 
This emphasizes
the differences between a quantum model with $S=1$ and the
classical limit $S\gg 1$ investigated by Shen {\it et al.}~\cite{ShenXie}
where fluctuations in the antiferromagnetic bonds completely vanish. 

An alternating ferromagnetic-antiferromagnetic spin exchange 
should also show up in the
spin-spin correlation function $\langle S^z_0 S^z_r \rangle$ showing 4
site periodicity (i.e. $\pi/2$-oscillations) and in a spin dimer correlation
function $\langle S^z_0 S^z_1 S^z_r S^z_{r+1} \rangle$
with long range order and $\pi$-oscillations. 
\begin{figure}[!ht]
\includegraphics*[width=0.9\columnwidth]{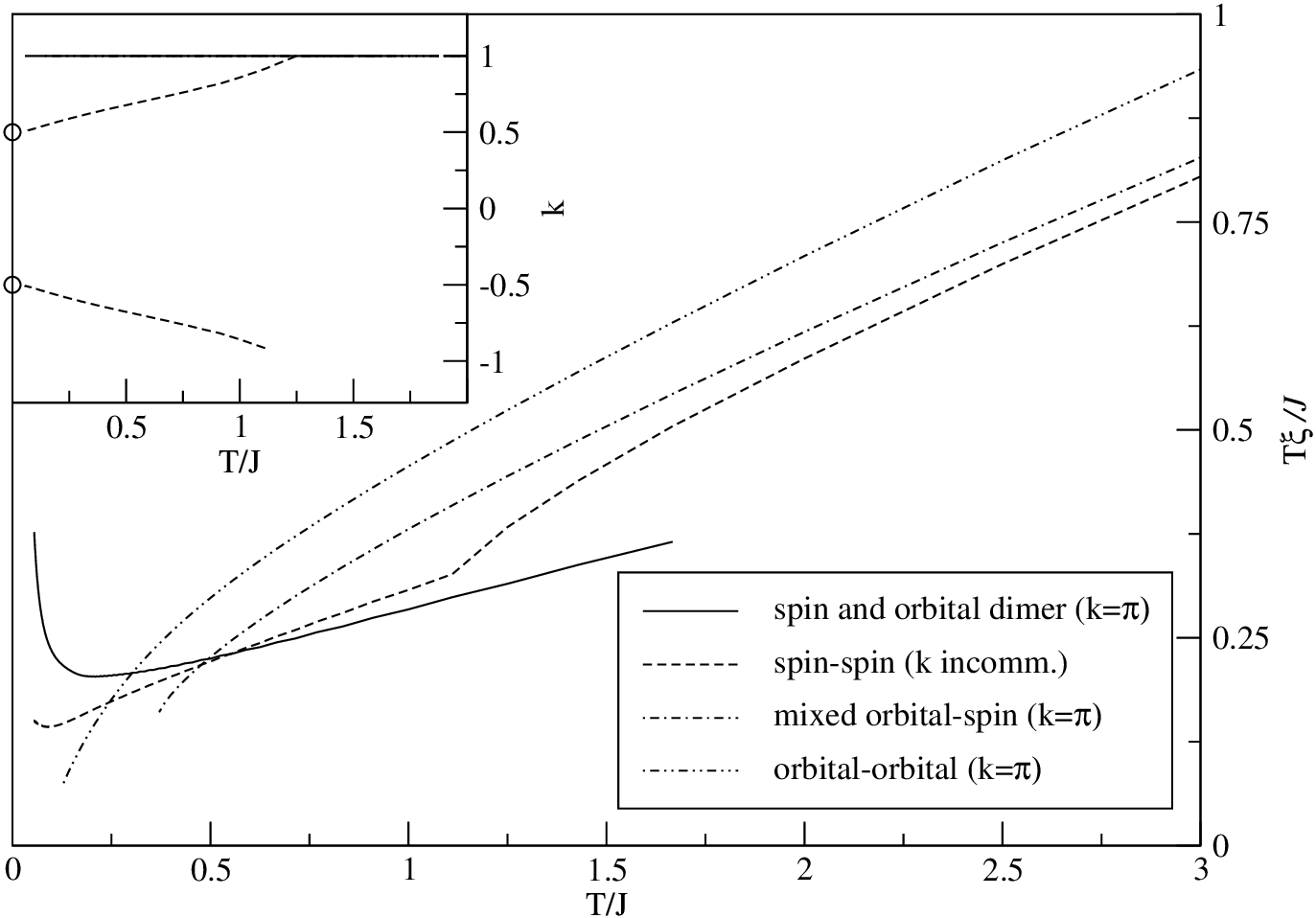}
\caption{The main figure shows the leading CLs times temperature 
in the isotropic case; the inset shows the corresponding wavevectors (in units of $\pi$) with circles denoting $k=\pm\pi/2$.}
\label{fig2} 
\end{figure}
In Fig.~\ref{fig2} numerical results for some leading CLs and the 
corresponding wavevectors are
shown. Because it is known from conformal field theory 
that CLs $\xi$ belonging to critical
excitations diverge as $\xi\sim 1/T$ in the low-temperature
limit, CL times temperature versus temperature is plotted, thus separating
correlation functions with long range order at $T=0$ from short 
range correlations. A dimer CL showing $k=\pi$ oscillations 
diverges as $\xi\sim \exp(\Delta_D/T)/\sqrt T$ 
with $\Delta_D\approx 0.08$ in the low-temperature limit indicating 
long-range dimer order at zero temperature. Second largest at 
low temperatures is a spin-spin CL showing
incommensurate oscillations approaching $\pi/2$ at zero temperature. Thus,
the numerical results are consistent with the picture above. 
Note also, that at high temperatures the orbitals have a short 
range antiferromagnetic order
and that a spin-orbital CL shows up, indicating the importance of 
mixed excitations.    

\begin{figure}[!ht]
\includegraphics*[width=0.9\columnwidth]{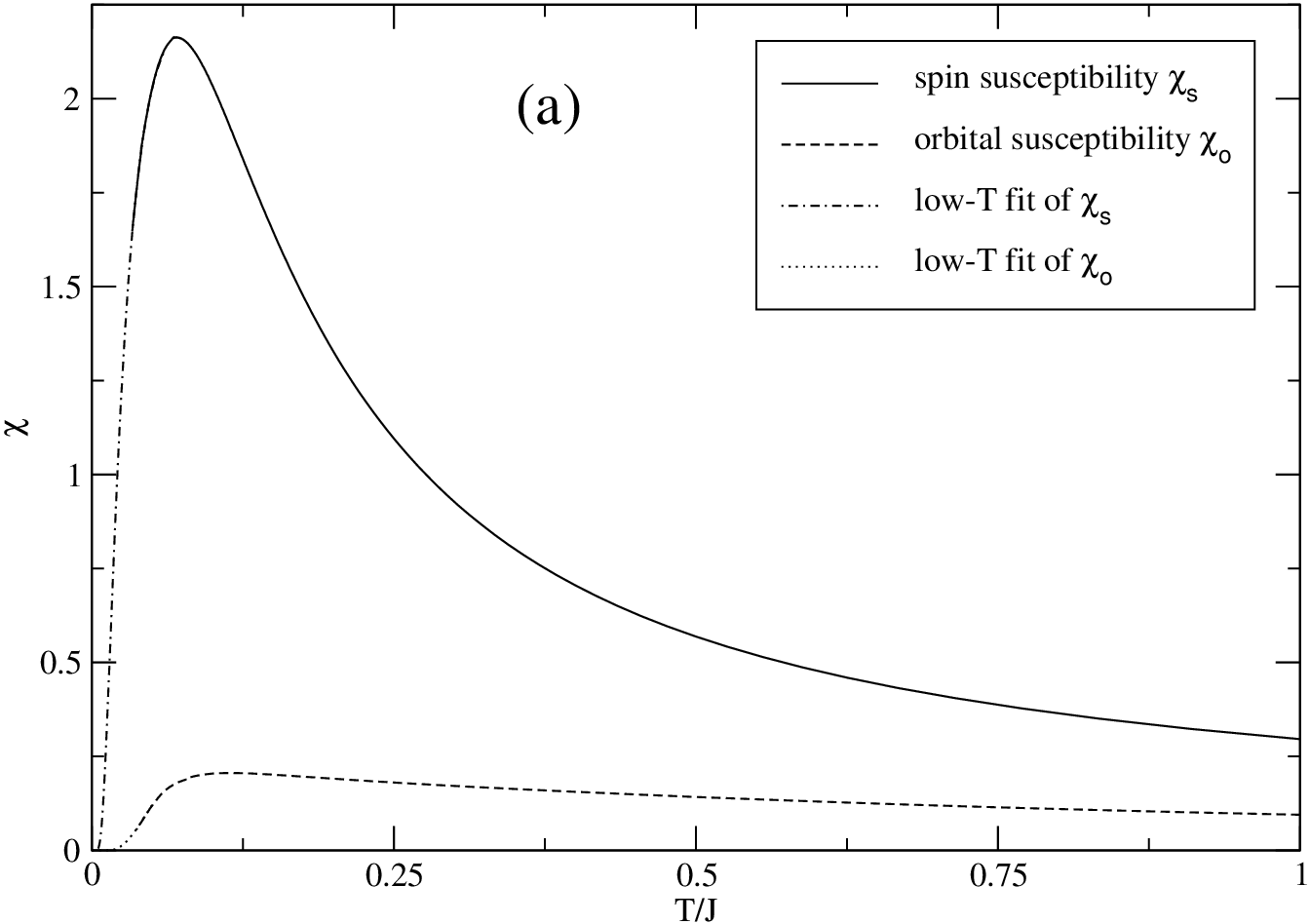}
\includegraphics*[width=0.9\columnwidth]{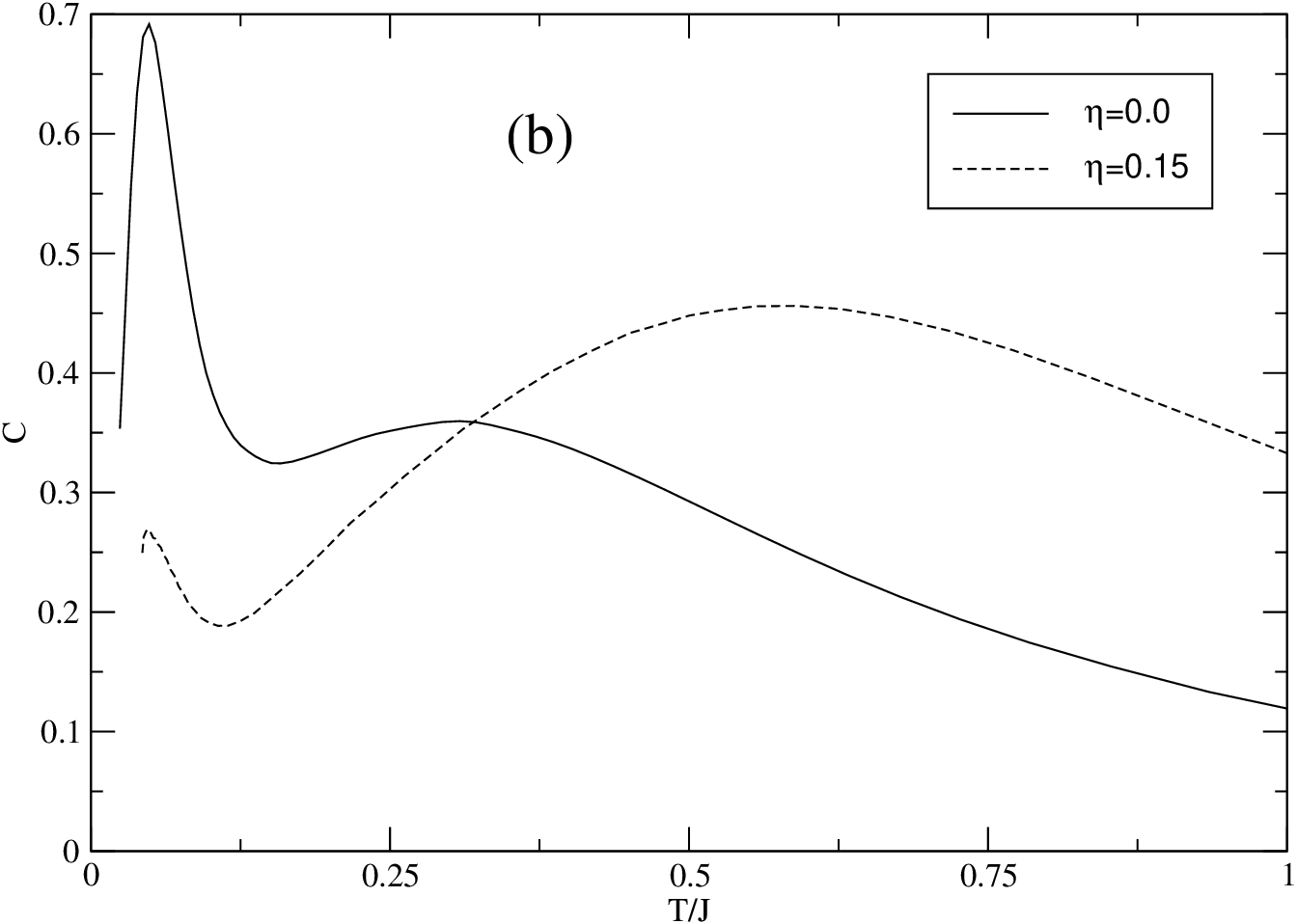}
\caption{(a) Spin and orbital susceptibility for $\eta=0$ with
low-temperature fits as specified in the text. (b) Specific heat for
the isotropic and the finite Hund's coupling case 
.}
\label{fig3}
\end{figure}

For a dimerized orbital pseudospin-$1/2$ as well as for an $S=1$
ferromagnetic-antiferromagnetic alternating Heisenberg chain a gap
in the excitation spectrum is expected as visible in the 
susceptibility data shown in Fig.~\ref{fig3}a. 
The quadratic dispersion of a gapped system leads in 1D to the 
low-temperature asymptotics $\chi \sim \exp (-\Delta /T) / \sqrt T$. 
Using this function for a fit, a spin gap $\Delta_s = (0.041\pm0.002)J$ 
and an orbital gap $\Delta_{orb} = (0.120 \pm 0.008)J$ is obtained.
Note that the orbital gap is
much smaller than the singlet-triplet splitting $
J$ of an isolated orbital dimer.  The reason is that in the coupled
spin-orbital system the spins form a singlet if the orbitals are excited to a
triplet configuration and thus, the much smaller gap of the orbital
triplet/spin singlet excitation appears in the orbital susceptibility data.
More precisely, the thermal gap visible here is half of the spectral gap
\footnote{The origin of the factor $1/2$ can be understood as follows: The
  elementary excitations are orbital "spinons" each with a gap $\Delta$, but
  they always appear in pairs so that $\Delta_{orb}^{spectr.}=2\Delta$. In the
  thermodynamic data the activated behaviour is characterized by the energy
  $\Delta$ of an elementary excitation irrespective of whether these
  excitations appear in pairs or not.}, so that we obtain
$\Delta_{orb}^{spectr.}\sim 0.24\, J$. This is very close to the value $J/4$
for an isolated bond, showing that the orbitals are nearly perfect dimers in
the ground state.
Dynamical coupling of two
sectors affects also the spin gap value. In a simple picture of
spin-$2$ objects coupled antiferromagnetically with $J/32$
~\cite{ShenXie}, one would expect
%
the $S=2$ Haldane gap of about $0.04(J/32)=J/800$ \cite{Yamamoto}.
The observed spin gap is in fact much larger due to the coupling 
between spins and orbitals: 
A spin excitation introduces additional antiferromagnetic couplings 
between the orbital singlets, thus raising also
the energy of the orbital sector without destroying the dimer state. 



Now we want to discuss the effects of finite Hund's coupling. 
As easily understood from Eqs. (\ref{eq1},\ref{eq2}) an additional, orbital free term $R$ 
tends to stabilize a ferromagnetic spin order
at large values of $\eta$. Numerical data for $\eta=0.15$
indeed show that $\langle \vec{S}_i\cdot\vec{S}_{i+1} \rangle$ is monotonously
increasing and reaches 1, the value expected for an uniform ferrochain, in 
the zero temperature limit (Fig.~\ref{fig1}).
This leads to an exact cancellation of the orbital anisotropy 
terms in the ground state (see Eqs.~\ref{eq1},\ref{eq2}) and therefore to an isotropic 
antiferromagnetic coupling of the orbitals.  
From the Bethe ansatz result for the 
Heisenberg chain one expects
$\langle\vec{\tau}_i\cdot\vec{\tau}_{i+1}\rangle=-\ln 2+1/4$ consistent with the
numerical result (see Fig.~\ref{fig1}). 
Low temperature asymptotics of the spin and orbital expectation 
values suggest
a first order phase transition between  
two possible ground states: Spin/orbital dimer phase and 
spin saturated ferrophase. A critical coupling $\eta_c\sim 0.11$
separating them at zero temperature 
is just slightly below realistic values for vanadium oxides 
$\sim 0.12$ \cite{Mizokawa}.

\begin{figure}[!ht]
\includegraphics*[width=0.9\columnwidth]{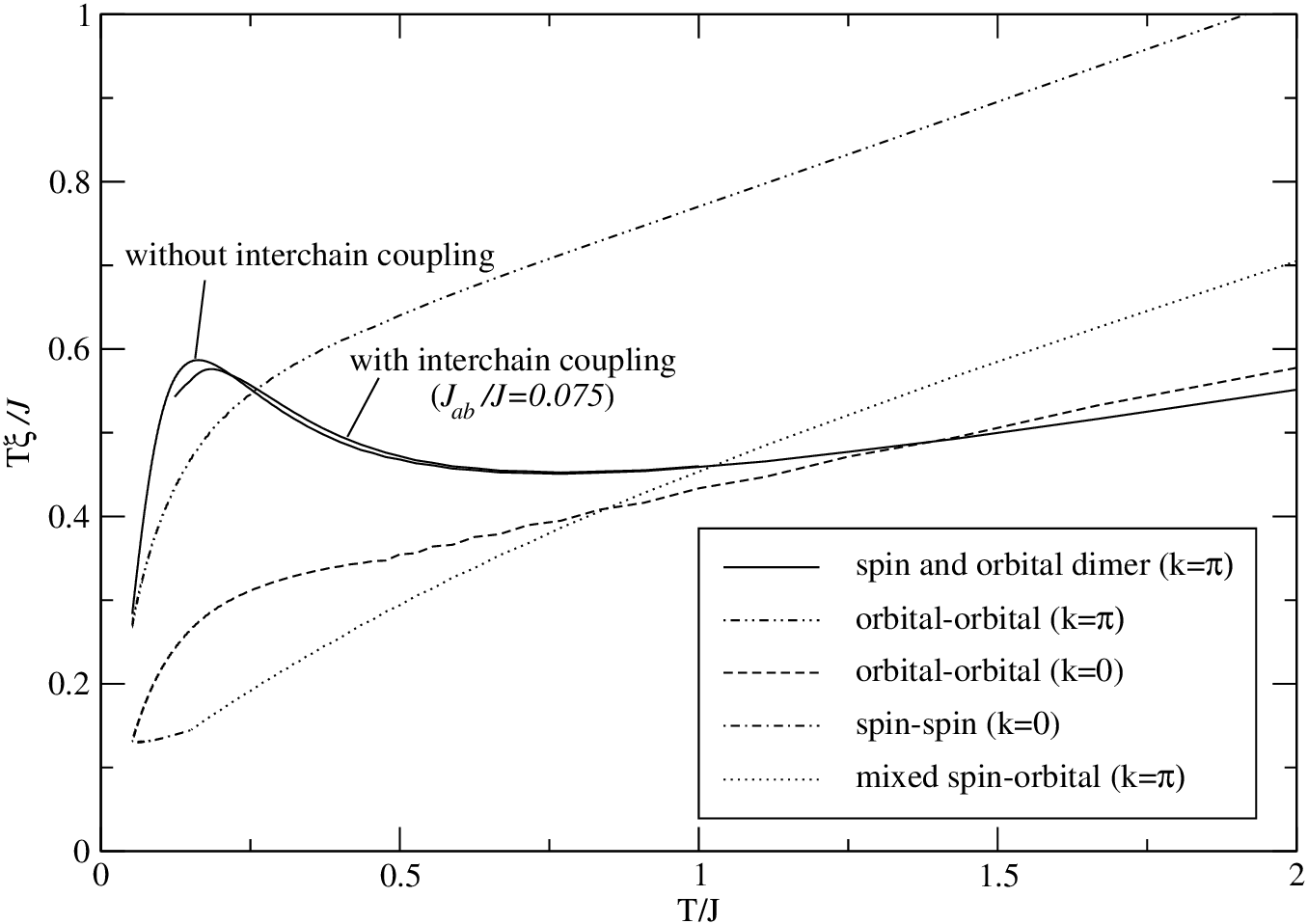}
\caption{The leading CLs times temperature for $\eta=0.15$. Note, that
all shown CLs have commensurate oscillations over the entire
temperature range.}
\label{fig6}
\end{figure}

The specific heat 
(Fig.~\ref{fig3}b) shows a two peak structure. 
The high temperature hump is due to orbital and mixed spin-orbital
excitations whereas the low-$T$ peak is of spin origin.
With increasing Hund's coupling,
the hump is shifted to higher temperatures, 
whereas the low-$T$ peak is smaller and shifted to
lower temperatures.  It is striking that the high-$T$ hump contains
such a large entropy weight. If it is due to orbital excitations only
its entropy weight would be $\ln 2$ but in fact it is approximately
1. This could be explained by a formation of an intermediate dimer
state at finite temperature. In this case the entropy weight of the
hump is given by the total entropy $\ln 6$ minus a spin entropy of
$S=2$ dimers given by $1/2\ln 5$ consistent with the numerical
data. To support this scenario of a temperature driven dimerization
further, we have calculated again the leading CLs (Fig.~\ref{fig6}).
The dimer CL with $\pi$-oscillations is 
indeed the largest CL in a
certain temperature range and hence, dimerization is in this
temperature range the leading instability towards the onset of long
range order in a 3D system.
Note also, a large orbital CL with $\pi$-oscillations indicating antiferromagnetic orbital
order. 
At low temperatures there is a non-oscillating spin CL present as expected for an uniform ferrochain, however, this 
ferromagnetic CL is strongly suppressed at finite temperature
due to a disorder effect of low-lying dimerized states. The competition
between uniform and dimerized states is evident also from Fig.~\ref{fig1}:
In comparison to the spin only case, where $J_{i,j}$ is fixed by its zero
temperature expectation value for $\eta=0.15$ and therefore all dimerization
effects are omitted, $\langle \vec{S}_i\cdot\vec{S}_{i+1} \rangle$ shows an unusual rapid decay of local spin correlations with temperature. 

Finally, we address the relevance of these findings to the intermediate
phase of YVO$_3$. Obviously we have to take the coupling between
different $c$-axis chains into account. The quasi 1D character of
orbital physics is supported by the $C$-type spin structure leading to
a weak interchain orbital coupling proportional to $\eta$ only
\cite{KhaliullinHorsch}. Interchain spin couplings can be explicitly
included in the numerical calculations as a mean field,
$4J_{ab}\langle S^z\rangle$. An estimation for $J_{ab}$ can be obtained
in the following way: Within linear response the
susceptibility of the 3D system is given by
$\chi_{3D}=\chi_{1D}/(1-4J_{ab}\chi_{1D})$.
Taking $T_{N1}/J\sim 0.25$ 
from experiment (a representative value of $J\sim 40\,$meV is used), 
demanding that $\chi_{3D}$ diverges at this
temperature and using the numerical data for $\chi_{1D}$ 
calculated at $\eta=0.15$, we find
$J_{ab}/J\sim 0.075$ in good agreement with the experimentally
observed value \cite{Keimer02}. In Fig.~\ref{fig6} the dimer
correlation length 
with this interchain molecular
field included is displayed, showing that the results above are only
weakly affected. The temperature range where dimerization is the
leading instability approximately coincides with the $C$-phase of
YVO$_3$. We therefore believe that the model explains why in YVO$_3$ a
$C$-type magnetic order is stable as an intermediate phase. The gain
of entropy due to the dimerization lowers the free energy $F=\langle H
\rangle-TS$ at finite temperature and stabilizes a formation of
alternating weak and strong ferrobonds along one cubic axis (e.g. $c$
axis). In other words, a dimerization of the {\it ferromagnetic} spin
chain occurs due to the {\it orbital Peierls effect}, in which thermal
spin fluctuations play a role of lattice degrees of freedom.
The effect is dynamical in the present 1D case.
However, as discussed before a weak but finite coupling between
different $c$-chains is present in YVO$_3$ which does not destroy
the quasi 1D spin-orbital physics but leads to the onset of long range
dimer order.  Doubling of the magnetic unit cell 
along the $c$ axis is then expected. In fact, a modulation of
ferrocouplings and related splitting of the magnon branches
has recently been observed by Ulrich {\it et al.}~\cite{Keimer02} 
in the $C$-phase of YVO$_3$.
As the dimerization is of electronic origin, 
concomitant lattice distortions are 
expected to be small because $t_{2g}$ orbitals are not bond oriented 
and only weakly coupled to the lattice.
%

Summarizing, the one-dimensional spin-orbital model with $S=1$ shows 
an intrinsic tendency towards dimerization. 
Large Hund's coupling can stabilize a uniform ferromagnetic ground state 
but strong dimer correlations develop again at finite temperature 
due to the orbital Peierls effect. Dimer states modulate spin and orbital
bonds and release high entropy. 
The quasi 1D physics is not destroyed when realistic interchain couplings
are taken into account on a mean field level.
The temperature driven
dimerization in the 
1D spin-orbital model could explain
the stability of the intermediate $C$-type phase in YVO$_3$.

\begin{acknowledgments}
The authors acknowledge valuable discussions with A.~Kl\"umper, B.~Keimer, A.M.~Ole\'s and P.~Horsch.   
J.S. acknowledges support by the DFG in SP1073.
\end{acknowledgments}

\end{document}